# Complex impedance and Raman spectroscopy of $Na_{0.5}(Bi_{1-x}Dy_x)_{0.5}TiO_3$ ceramics

Manal Benyoussef[a,*], Moneim Zannen[b], Jamal Belhadi[a], Bouchaib Manoun[c], Jean-Luc Dellis[a], Abdelilah Lahmar[a,*], Mimoun El Marssi[a],

[a] Laboratory of Physics of Condensed Matter (LPMC), University of Picardie Jules Verne, Scientific Pole, 33 rue Saint-Leu, 80039 Amiens Cedex 1, France.

[b] Laboratory of Interfaces and Advanced Materials (LIMA), Faculty of Sciences of Monastir (University of Monastir), Bd. Of the Environnement, Monastir 5019, Tunisia.

[c] Univ Hassan 1[er] Laboratoire des Sciences des Matériaux, des Milieux et de la Modélisation (LS3M), 25000, Khouribga, Morocco.

**Abstract**

In this work structural refinement, complex impedance spectroscopy (CIS) and Raman spectroscopy have been investigated on $Na_{0.5}(Bi_{1-x}Dy_x)_{0.5}TiO_3$ (xDyNBT) ceramic systems with x = 0, 2, 5 and 15%. The pure NBT, 2DyNBT and 5DyNBT compounds crystallize in a rhombohedral *R3c* structure while the 15DyNBT composition crystallizes in an orthorhombic *Pnma* structure. We reported that dysprosium addition affects the phase transition temperatures as well as the dielectric losses. The electrical transport at high temperatures was investigated using the CIS over a wide frequency range ($10 - 10^6$ Hz). The studied samples showed a non-Debye type process, with a short-range relaxation for the pure NBT and a coexistence of both localized and long-range relaxations of charge carriers for the 2DyNBT and 5DyNBT compounds. For the high concentration, 15DyNBT, a short-range relaxation is observed. Moreover, using a brick-layer model we discuss the resistance and capacitance of the different contributors (grain and grain boundaries) in our samples. High temperature Raman spectroscopy investigation was performed in order to follow the temperature evolution of the structural transformations on ferroelectric compounds. Anomalies in the temperature evolution of the vibrational modes are seen to correlate well with the temperature transitions observed from dielectric measurements.

**Corresponding authors:** Manal Benyoussef: manalbenyoussef@gmail.com

Abdelilah Lahmar: abdel.ilah.lahmar@u-picardie.fr

## 1. Introduction

In the recent years, the environment concern has grown in importance due to climate change and global warming. For this reason, it was strongly advised to develop non-toxic materials to replace lead-based ferroelectric materials in various devices such as actuators, transducers, filters, multilayer capacitors, etc. [1]. Bismuth is similar to lead in mass and outermost electron shell, and thus could replace it in ferroelectric materials [2]. In 1960, Smolensky *et al.* [3] discovered for the first time the ferroelectric $Na_{0.5}Bi_{0.5}TiO_3$ (NBT) material, which is now the most promising candidate to replace lead-based systems such as $Pb(Zr,Ti)O_3$ systems known by their excellent piezoelectric properties in the morphotropic phase boundary vicinity [4,5].

Besides, the ionic conductivity of NBT at high temperatures was considered differently depending on the target application. In fact, owing to its high conductivity NBT is known to be an excellent oxide-ion conductor for technical applications in electrochemical devices and sensors [6–8]. For instance, Shih *et al.* revealed an enhanced conductivity for lithium doped NBT system [8]. Conversely, this conductivity is seen as problematic for piezoelectric, ferroelectric and for energy storage applications [9–12]. Numerous works were interested on rare earth doping NBT material in order to resolve those issues [13–15]. Rare earth doping has been proved to be a simple and elegant way to enhance the physical properties of the pure matrix [12,16,17]. In a recent work, we reported that a small dysprosium substitution in the A-site of NBT matrix led to a decrease of the coercive field and to a stabilization of the antiferroelectric like behavior at high temperatures which is interesting for energy storage capacitors operating at high temperatures [12]. Note that dysprosium concentrations higher than 10%, were reported to have a paraelectric behavior. The considerable difference in the ionic radii between bismuth ($r_{Bi3+}$ = 1.03Å) and dysprosium ($r_{Dy3+}$ = 0.912Å) explains the change from ferroelectric to paraelectric like behavior at the told concentration. The DyNBT with high concentrations are interesting for applications in the view to replace the C0G capacitors. Detailed structural refinement was done for NBT parent compound, as well as for different $Dy^{3+}$ doping level in order to confirm the occurring structural change.

It is worthy to mention that, the control of the electrical transport properties as well as the understanding of its mechanism in particularly at high temperatures should be considered before any application. The impedance spectroscopy is one of the commonly used technique to dept the electrical properties in ferroelectric materials. In fact, by analyzing the temperature and frequency evolution of the impedance data, plenty information can be extracted such as the relaxation process, conduction mechanism, and can also give a correlation between the microstructure (grain or grain boundary effect) and the properties of ferroelectric materials. For these aims, dielectric and impedance analysis are conducted in the present work for the three ferroelectric xDyNBT compositions (x = 0, 2, 5%) and the paraelectric 15DyNBT compound to better understand the electrical behavior, especially at high operating temperatures where the technological applications are expected.

Moreover, NBT is also known for its manifold dielectric behaviors. Two significant temperatures can be extracted from the dielectric investigations of NBT system: the temperature of the dielectric permittivity maximum ($T_m$) which describes the antiferroelectric like to paraelectric phase transition, and the depolarization temperature ($T_d$) which defines the ferroelectric to antiferroelectric like phase transition. Nevertheless, the nature of this last is still controversial, due to the fact that no long-range order change has been revealed by structural investigations near $T_d$ [18]. Thus, the term relaxor antiferroelectric have been employed by Ma *et al.* to describe this behavior [19]. Knowing that ferroelectric relaxors contain FE nanodomains/polar nanoregions dispersed in a cubic matrix, the antiferroelectric relaxors could be defined as the antiparallel displacement of the A-site cations ($Na^+$, $Bi^{3+}$) toward B-site cations ($Ti^{4+}$). Such displacement occurs within each individual nanodomain, resulting in AFE-nanodomains. The investigation of the AFE like behavior is found to be more sensitive to methods probing the local structure and dynamics instead of conventional x-ray diffraction methods. Raman spectroscopy is known to be a powerful method for analyzing the mesoscopic-scale structure and phase transitions. It should be noticed that in NBT based systems, the ferroelectrically active $Bi^{3+}$ and $Ti^{4+}$ are the main responsible elements for driving A-site as well as B-site cations vibrations. One should highlight that the A-site cationic vibrations are expected in the low frequency part of the Raman spectra (<150cm$^{-1}$). Therefore, to strengthen the structural refinement and dielectric studies, advanced Raman investigation is conducted from 10 cm$^{-1}$ to 1000 cm$^{-1}$ within a temperature range from room temperature to 580°C for the ferroelectric compositions with rhombohedral *R3c* structure (x= 0, 2, and 5%). Moreover, a comprehensive analysis was performed over the entire spectral range in order to examine the effect of temperature and dysprosium introduction on the vibrational modes and phase transitions.

## 2. Experimental

The xDyNBT ceramics were prepared using conventional powder processing method starting from high purity raw materials $Bi_2O_3$ (Merck 99%), $Na_2CO_3$ (Aldrich 99%), $TiO_2$ (Aldrich 99.9%) and $Dy_2O_3$ (Aldrich 99%). Preparation procedure is detailed elsewhere [20,21]. Room temperature powder X-ray diffraction patterns were recorded on a discover Bruker diffractometer (CuKα = 1.5406 Å). The electrical measurements were carried out using a Solartron Impedance analyzer SI-1260. The Raman spectra were recorded using a micro-Raman Renishaw spectrometer equipped with a CCD detector. The green laser was used to excite the samples (514.5 nm). Both stokes and anti-stokes Raman spectra were recorded in order to extract the local temperature of our compounds which was obtained from the ratio intensity of the stokes and anti-stokes part ($I_S/I_{AS}$) of the mode at ~280 cm$^{-1}$ for all temperatures. It was observed that for a laser power of 1 mW, the local temperatures were in agreement with the expected ones for the three compounds. In order to analyze the temperature evolution of Raman modes, we first subtract a linear base-line from our measured Raman spectra. Therefore, these last spectra were Bose–Einstein corrected in order to remove the temperature effect on the peak intensities. Therefore, a reduction using the occupation factor was done for all the spectra; $I_S = \frac{I_{measured}}{n(\omega,T)+1}$, with $n(\omega,T) = \frac{1}{e^{\frac{\hbar\omega}{kT}}-1}$; ω is the phonon wavenumber, T the temperature, ℏ the reduced Planck constant and k is the Boltzmann constant. Lorentzian functions were used to fit our spectra. A first fit was done at room temperature for our samples, and was therefore used as a starting model for the following temperatures.

## 3. Results and discussion

### 3.1. Structural investigations

Figure 1 (a) show the X-ray diffraction (XRD) patterns of xDyNBT (x = 0, 2, 5, and 15%) sintered ceramics. A stable and pure perovskite phase was detected for all compositions. The XRD reflections were indexed in the rhombohedral phase with *R3c* as a space group for the pure, 2DyNBT and 5DyNBT. The high concentration structure (15DyNBT) is indexed in the orthorhombic phase with *Pnma* as a space group.

Rietveld refinement of the XRD data of the materials was used for further structural characterizations. The full patterns refinements were carried out by means of the Rietveld method using the FullProf program [22] integrated in WinPlotr software [23]. The peak shape was described by a pseudo–Voigt function. In addition, the background level was modelled using 6 coefficients polynomial function. The refined parameters were: scale factor, fractional atomic coordinates (*x*, *y*, *z*), occupancy rate of atoms, isotropic displacement parameter in Å$^2$, preferred orientation parameters, overall isotropic displacement (temperature), zero–point of detector, cell parameters (*a*, *b*, *c*, *α*, *β*, *γ*), refine background with 6 coefficients polynomial function, half–width parameters (*u*, *v*, *w*), profile shape parameter (*η*) and asymmetry parameters.

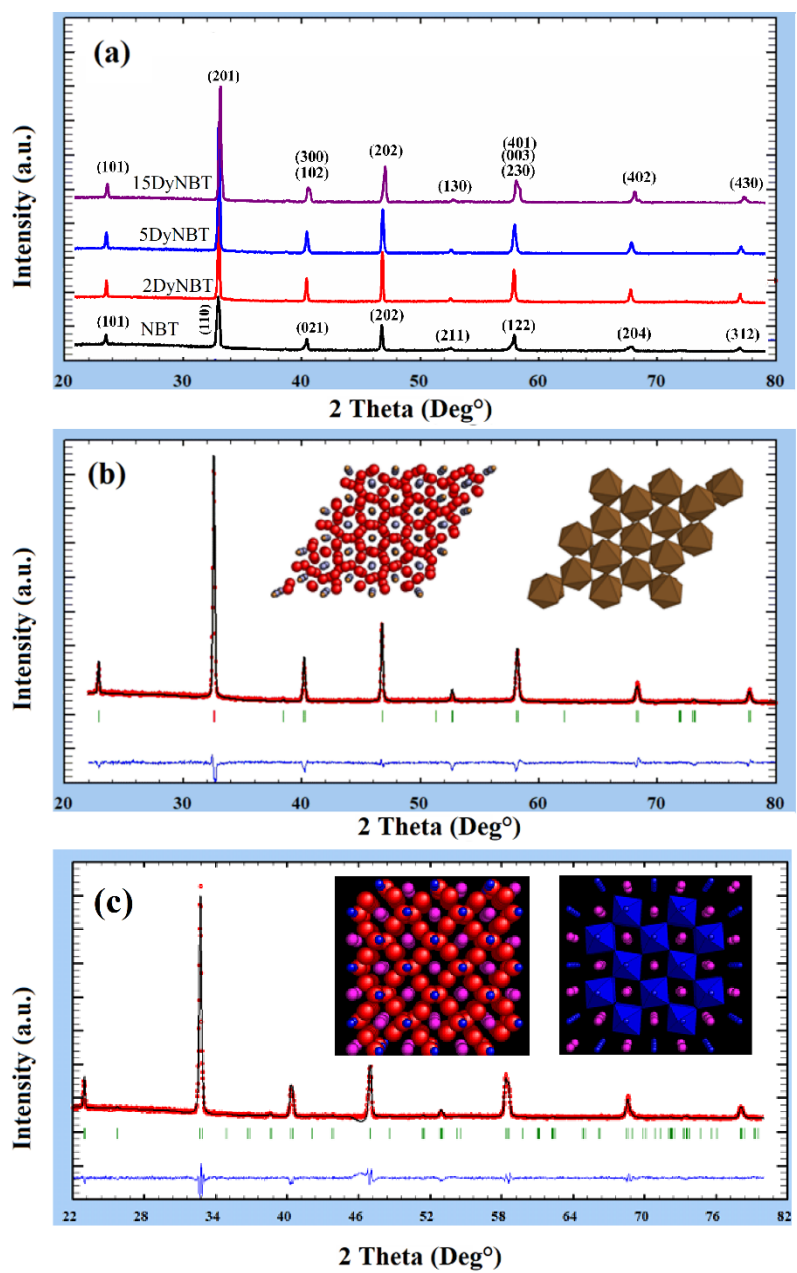

**Figure 1**: **(a)** X-ray diffraction patterns of all xDyNBT (x=0, 2, 5 and 15%) sintered ceramics. Rietveld refinement X-ray powder diffraction pattern of **(b)** 5DyNBT and **(c)** 15DyNBT: experimental, calculated and their difference. The projection views of the structures are shown in the inset.

For the pure NBT, 2DyNBT and 5DyNBT compositions, the refinements were carried out assuming a rhombohedral symmetry (*R3c*). In this model $Na^+$, $Bi^{3+}$ and $Dy^{3+}$ are localized at the *6a* sites (0, 0, z), $Ti^{4+}$ cations are occupying also the *6a* positions (0, 0, z) and the oxygen atoms occupy the *18b* general positions (x, y, z). Concerning the refinement of the higher composition (15DyNBT), it was carried out assuming an orthorhombic symmetry (*Pnma*). In this model Na$^+$, Bi$^{3+}$ and Dy$^{3+}$ are localized at the *4c* sites (x, ¼, z), Ti$^{4+}$ cations are occupying also the *4b* positions (1/2, 0, 0) and the oxygen atoms occupy both 4c sites and the *8d* general positions (x, y, z).

**Table 1:** Details of Rietveld refinement parameters for NBT, 2DyNBT, 5DyNBT and 15DyNBT compositions.

| Composition | **Pure NBT** | **2DyNBT** | **5DyNBT** | **15DyNBT** |
|---|---|---|---|---|
| Wavelength (Å) | λKα1 = 1.5406 | λKα1 = 1.5406 | λKα1 = 1.5406 | λKα1 = 1.5406 |
| Step scan increment (°2θ) | 0.026 | 0.026 | 0.026 | 0.026 |
| 2θ range (°) | 20-80 | 20-80 | 20-80 | 20-80 |
| Program | FULLPROF | FULLPROF | FULLPROF | FULLPROF |
| Zero point (°2θ) | 0.055 (7) | -0.02 (1) | 0.003 (6) | -0.027 (4) |
| Pseudo-Voigt function PV = η L + (1 − η) G | η= 0.06(1) | η= 0.13(1) | η= 0.15(6) | η= 0.26(2) |
| Caglioti parameters | U= 0.23 (7) V= -0.13(3) W= 0.032 (5) | U= 0.12 (4) V= -0.06(3) W= 0.024 (4) | U= 0.17 (4) V= -0.006(3) W= 0.033 (6) | U= 0.058 (3) V= -0.023(2) W= 0.037 (3) |
| No. of reflections | 40 | 40 | 40 | 121 |
| No. of refined parameters | 24 | 24 | 24 | 28 |
| Space group | *R3c* | *R3c* | *R3c* | *Pnma* |
| a (Å) b (Å) | 5.488(2) | 5.485(1) | 5.483(1) | 5.445(1) 7.736(2) |
| c (Å) | 13.508(3) | 13.468(2) | 13.467(3) | 5.479(2) |
| RF | 9.5 | 5.5 | 8.2 | 7.5 |
| RB | 11.4 | 5.5 | 7.2 | 8.1 |
| Rp | 7.7 | 6.8 | 8.6 | 8.0 |
| Rwp | 10.3 | 9.4 | 11.5 | 11.7 |

As an example, figure 1 (b) and (c) illustrates the typical Rietveld refinement patterns along with the difference plot at room temperature for 5DyNBT and 15DyNBT sintered ceramic, respectively.

Relatively low residuals of the refinements have been obtained (Table 1). For all compositions, the refined atomic occupations are listed in Table 2. Ti$^{4+}$ cations reside in octahedral environment; Ti-O distances show short and long lengths leading to distorted octahedra, the average value of Ti$^{4+}$-O is around 1.95 Å. These octahedra are alternatively connected and extended in three dimensions. The analysis of various inter-atomic distances (Table 3) shows that Na/Bi/Dy atoms form (Na/Bi/Dy)$O_{12}$ polyhedra, the Na/Bi/Dy–O bond lengths average value is calculated to be around 2.75 Å.

**Table 2:** Refined structural parameters of NBT, 2DyNBT, 5DyNBT and 15DyNBT compositions.

| Atom | Site | x | y | z | B(Å) | Occupancy |
|---|---|---|---|---|---|---|
| **NBT** | | | | | | |
| Na/Bi | 6a | 0 | 0 | 0.2627(4) | 0.83(6) | 0.5/0.5 |
| Ti | 6a | 0 | 0 | 0.0063 (6) | 0.52(4) | 1 |
| O | 18b | 0.1263(9) | 0.3369(4) | 0.0834(5) | 1.83(7) | 1 |
| **2DyNBT** | | | | | | |
| Na/Bi/Dy | 6a | 0 | 0 | 0.2623(4) | 0.39(4) | 0.5/0.49/0.01 |
| Ti | 6a | 0 | 0 | 0.0059 (6) | 0.27(3) | 1 |
| O | 18b | 0.1258(6) | 0.3363(3) | 0.0829(5) | 1.61(5) | 1 |
| **5DyNBT** | | | | | | |
| Na/Bi/Dy | 6a | 0 | 0 | 0.2589(3) | 0.40(4) | 0.5/0.475/0.025 |
| Ti | 6a | 0 | 0 | 0.0059 (4) | 0.27(3) | 1 |
| O | 18b | 0.1270(5) | 0.3340(4) | 0.0831(5) | 1.60(6) | 1 |
| **15DyNBT** | | | | | | |
| Na/Bi/Dy | 4c | 0.0018(5) | 0.25 | 0.0139(9) | 0.47(5) | 0.5/0.425/0.075 |
| Ti | 4b | 0.5 | 0 | 0 | 0.34(4) | 1 |
| O | 4c | 0.4156(3) | 0.25 | 0.0037(4) | 1.52(7) | 1 |

**Table 3:** Selected inter-atomic distances (Å) and O-Ti-O angles for NBT, 2DyNBT, 5DyNBT and 15DyNBT compositions.

| Pure NBT | | | |
|---|---|---|---|
| 3×(Na/Bi)-O | 2.984 (7) | | |
| 3×(Na/Bi)-O | 2.919 (5) | 3×Ti-O | 1.924 (7) |
| 3×(Na/Bi)-O | 2.624 (8) | 3× Ti-O | 2.002 (9) |
| 3×(Na/Bi)-O | 2.523 (7) | <Ti-O> | 1.963 |
| <(Na/Bi)-O> | 2.761 | | |
| 3×O-Ti-O | 175.47(6) | 3×O-Ti-O | 86.98(9) |
| 3×O-Ti-O | 93.36(5) | 3×O-Ti-O | 89.82(9) |
| 3×O-Ti-O | 89.67(6) | | |

| 2DyNBT | | 5DyNBT | | 15DyNBT | |
|---|---|---|---|---|---|
| | | | | 1×(Na/Bi/Dy)-O1 | 3.193 (4) |
| | | | | 1×(Na/Bi/Dy)-O1 | 2.254 (3) |
| 3×(Na/Bi/Dy)-O | 2.978 (4) | 3×(Na/Bi/Dy)-O | 2.963 (3) | 1×(Na/Bi/Dy)-O1 | 2.874 (5) |
| 3×(Na/Bi/Dy)-O | 2.905 (3) | 3×(Na/Bi/Dy)-O | 2.860 (3) | 1×(Na/Bi/Dy)-O1 | 2.684 (6) |
| 3×(Na/Bi/Dy)-O | 2.611 (5) | 3×(Na/Bi/Dy)-O | 2.652 (6) | 2×(Na/Bi/Dy)-O2 | 2.878 (5) |
| 3×(Na/Bi/Dy)-O | 2.518 (6) | 3×(Na/Bi/Dy)-O | 2.525 (4) | 2×(Na/Bi/Dy)-O2 | 2.559 (4) |
| <(Na/Bi/Dy)-O> | 2.753 | <(Na/Bi/Dy)-O> | 2.750 | 2×(Na/Bi/Dy)-O2 | 2.919 (3) |
| | | | | 2×(Na/Bi/Dy)-O2 | 2.610 (4) |

|  |  |  |  | <(Na/Bi/Dy)-O> | 2.745 |
|---|---|---|---|---|---|
| 3×Ti-O<br>3× Ti-O<br><Ti-O> | 1.917(4)<br>1.995(6)<br>1.956 | 3×Ti-O<br>3× Ti-O<br><Ti-O> | 1.908(4)<br>2.0003(9)<br>1.954 | 2× Ti-O1<br>2× Ti-O2<br>2× Ti-O3<br><Ti-O> | 1.988(4)<br>2.035(6)<br>1.856(3)<br>1.96 |
| 3×O-Ti-O<br>3×O-Ti-O<br>3×O-Ti-O<br>3×O-Ti-O<br>3×O-Ti-O | 175.48(2)<br>93.51(3)<br>89.51(4)<br>87.14(6)<br>89.66(7) | 3×O-Ti-O<br>3×O-Ti-O<br>3×O-Ti-O<br>3×O-Ti-O<br>3×O-Ti-O | 175.80(3)<br>93.25(4)<br>89.62(3)<br>87.33(5)<br>89.65(6) | 1×O1-Ti-O1<br>2×O1-Ti-O2<br>2×O1-Ti-O2<br>2×O1-Ti-O2<br>2×O1-Ti-O2<br>2×O2-Ti-O2<br>2×O2-Ti-O2<br>2×O2-Ti-O2 | 179.97(2)<br>79.81(3)<br>81.80(4)<br>100.19(6)<br>98.20(7)<br>89.85(3)<br>180<br>90.15(2) |

*3.2. Dielectric investigations*

The temperature evolution of the permittivity for xDyNBT (x=0, 2, 5, 15%) ceramics are plotted in figure 2 (a-d) at different frequencies. In the case of the pure compound (x=0), three dielectric anomalies can be distinguished in accordance with the literature [24–26]. The low temperature anomaly ($T_d$ ~230°C) corresponds to the depolarization temperature associated to the ferroelectric to the antiferroelectric-like transition [27]. Remind that, NBT has a rhombohedral (*R3c*) structure from room temperature to 200°C, beyond this temperature the ferroelectric domains begin to disappear and a modulated structure is taking place between 230°C and 300°C in which *Pnma* orthorhombic sheets appear within the *R3c* matrix. Thus, the antiferroelectric behavior of NBT in this temperature range can be explained by the reorientation of the polar vector, since the cations are displaced along $[u0w]_p$ in the orthorhombic (*Pnma*) structure, which were initially oriented along the $[111]_p$ for the rhombohedral (*R3c*) structure [28]. Another anomaly is noticed in the three compounds around 300°C denoted as being $T_1$. Dorcet *et al.* clarify the origin of this anomaly using in-situ temperature dependent TEM study [28,29]. The appearance of new *½(oee)* superstructure reflections at 300°C, reveals a complete symmetry change from the modulated phase (200 - 300°C) to the orthorhombic (*Pnma*) phase. The last dielectric anomaly is found at $T_m$ ~ 340°C and is defining the orthorhombic (*Pnma*) to the tetragonal (*P4bm*) phase transition.

The introduction of dysprosium into NBT matrix leads to some changes in the dielectric behaviors. We first observe a decrease in the dielectric permittivity values from 2700 (NBT) to 440 (15DyNBT). In addition, we noticed that the $\varepsilon_r$ (T) curve near $T_m$ become broader as a function of the increase of $Dy^{3+}$ concentration together with a slight shift to higher temperatures for the low doping. For 15DyNBT (figure 2 (d)), the anomalies can't be distinguished due to the very broad $\varepsilon_r$(T) curve and the decrease of the permittivity as a function of temperature.

The broadening of this anomaly as well as the decrease of $\varepsilon_r$ values can be related to the increase of A-site disorder in NBT matrix. Concerning the bump presenting $T_1$ it becomes easily observable for 2% of $Dy^{3+}$ doping composition.

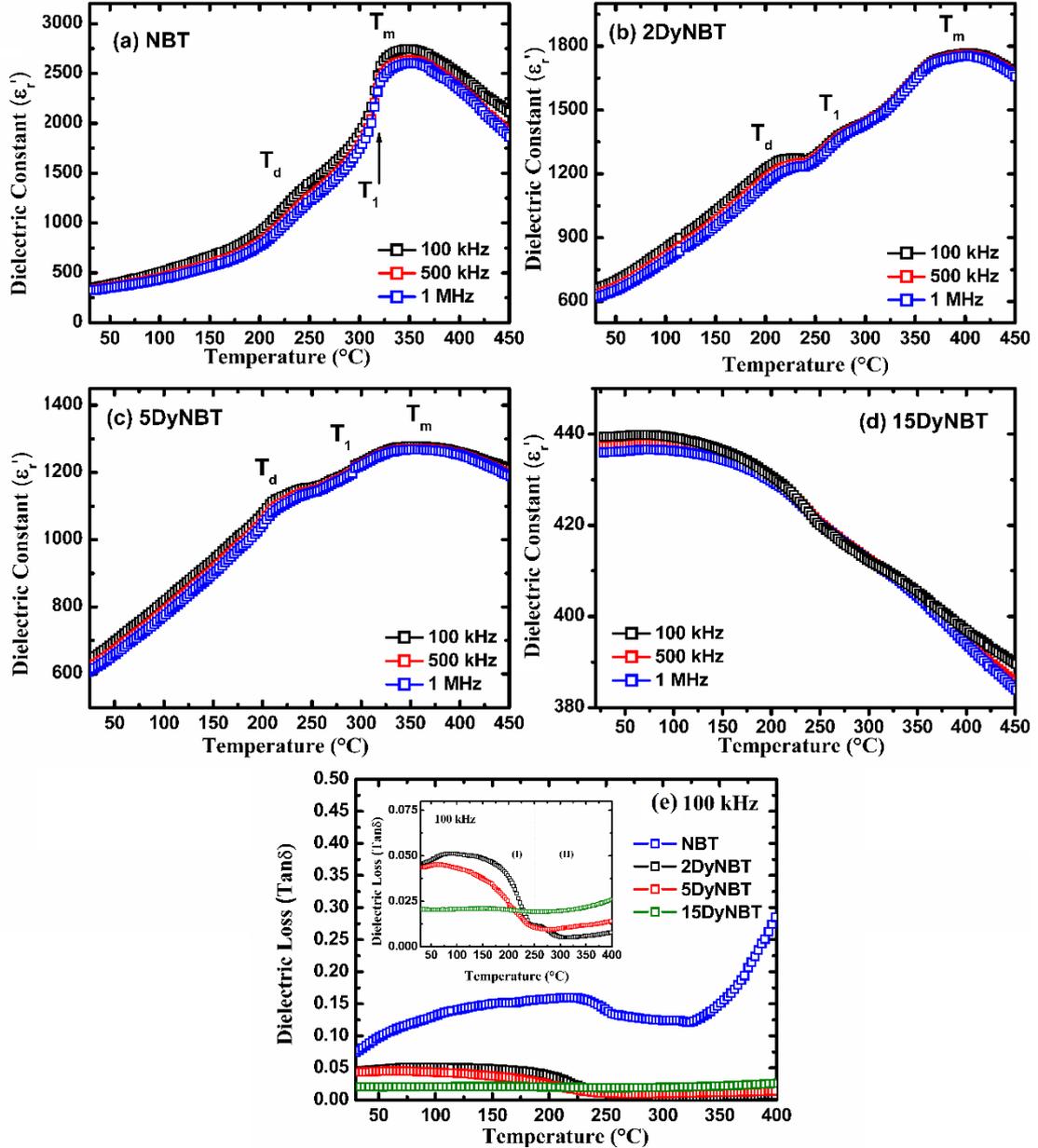

**Figure 2:** Dielectric constant of **(a)** NBT, **(b)** 2DyNBT, **(c)** 5DyNBT, and **(d)** 15DyNBT at three different frequencies. **(e)** Dielectric losses at 100 kHz.

This last observation is due to the shift of both $T_1$ and $T_m$ anomalies to lower and higher temperatures, respectively. Conversely, NBT and 5DyNBT are characterized by an overlapping of $T_m$ and $T_1$ anomalies. Besides, the depolarization temperature is getting more pronounced as a function of doping with a shift to lower temperatures from ~230°C to ~200°C for 2DyNBT and 5DyNBT. Notice that the dielectric permittivity values at $T_d$ didn't decrease significantly (1300 for NBT, 1100 for 5DyNBT), in contrary to the drastic decrease of $\varepsilon_r$ at $T_m$. Accordingly, for 5DyNBT both anomalies have very close $\varepsilon_r$ values with $T_d$ being more pronounced than the very broad $T_m$ anomaly.

Figure 2 (e) presents the dielectric losses of the pure and doped compounds at 100kHz. A huge decrease of the dielectric losses is observed for the doped compounds compared to the pure. We notice that the behavior of the dielectric losses can be divided into two regions for all compounds; from room temperature to 250°C (I) the tan δ curve is concave when from 250°C to the high temperatures (II) they are convex (inset of figure 2(e)). It can be observed that the dielectric losses decrease as a function of the increase of doping for the region I, in contrary to the region II where an inverse behavior is noticed. Moreover, 2DyNBT and 5DyNBT compositions have lower dielectric losses in the high temperature region (II) compared to the first temperature region (I) which is very interesting for high temperature applications.

*3.3. Complex impedance spectroscopy (CIS)*

The CIS is generally used in order to analyze the electrical response in a wide temperature and frequency range [30]. It is a well-used method for the study of the electrical properties and the microstructure of materials. The impedance measurements on a given material will give us both the resistive and the reactive components data corresponding to the real and imaginary parts, respectively. It can be displayed conventionally in a complex plane plot (Nyquist diagram) in different representations; Complex permittivity $\varepsilon(\omega) = \varepsilon' - j\varepsilon''$, complex impedance $Z(\omega) = Z' - Z''$, complex admittance $Y(\omega) = Y' + Y'' = \frac{1}{Z(\omega)} = j\omega C_0 \varepsilon$, and complex modulus $M(\omega) = M' + jM'' = \frac{1}{\varepsilon(\omega)} = j\omega C_0 Z$. Where the angular frequency is ω = 2πf, and $C_0$ is the vacuum capacitance.

Figure 3 (a-d) present the variation of the imaginary part of the complex impedance as a function of the frequency at different temperatures for all studied compounds. We can notice that Z″ increases in magnitude to reach a maximum value at a certain frequency called the relaxation frequency, after which the magnitude decreases. The relaxation frequency is depending on the studied compound and is observed to shift to higher frequencies for higher temperatures which indicates a thermally activated non-Debye-like process [31]. The departure from ideal Debye case is confirmed by the FWHM which is for all the studied samples greater than $log\frac{2+\sqrt{3}}{2-\sqrt{3}} = 1.14$ decades (FHWM for ideal Debye case) [32]. Furthermore, the introduction of dysprosium into NBT matrix seems to shift the relaxation to lower frequencies. As it can be seen in figure 3 (e), the comparison of $Z'$ versus frequency at 500°C for the pure and doped compounds confirms this behavior. Firstly, a plateau independent frequency is observed at low frequency, then a decrease occurs at the so-called relaxation frequency in $Z'$.

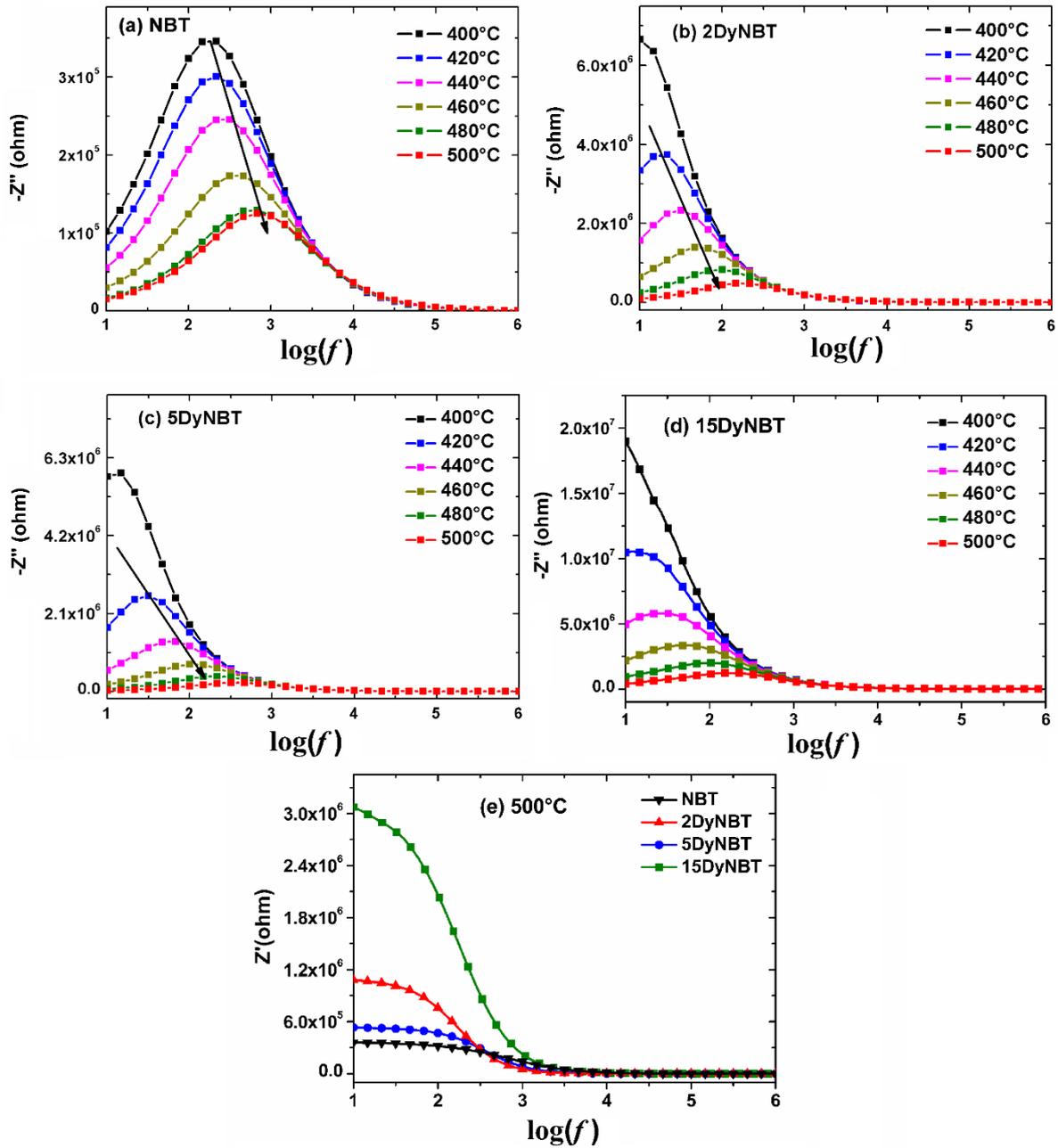

**Figure 3:** Variation of the imaginary part of complex impedance ($Z''$) with frequency **(a)** NBT, **(b)** 2DyNBT, **(c)** 5DyNBT, **(d)** 15DyNBT at different temperatures, and **(e)** variation of the real part of complex impedance ($Z'$) with frequency for all compounds at 500°C.

The study of the electric modulus (M*) is important, in order to observe clearly both grain and grain boundary contributions. In addition, it can give details about the electrical transport as well as the different types of polarization, due to the suppression of the electrode polarization effect in M* [33].

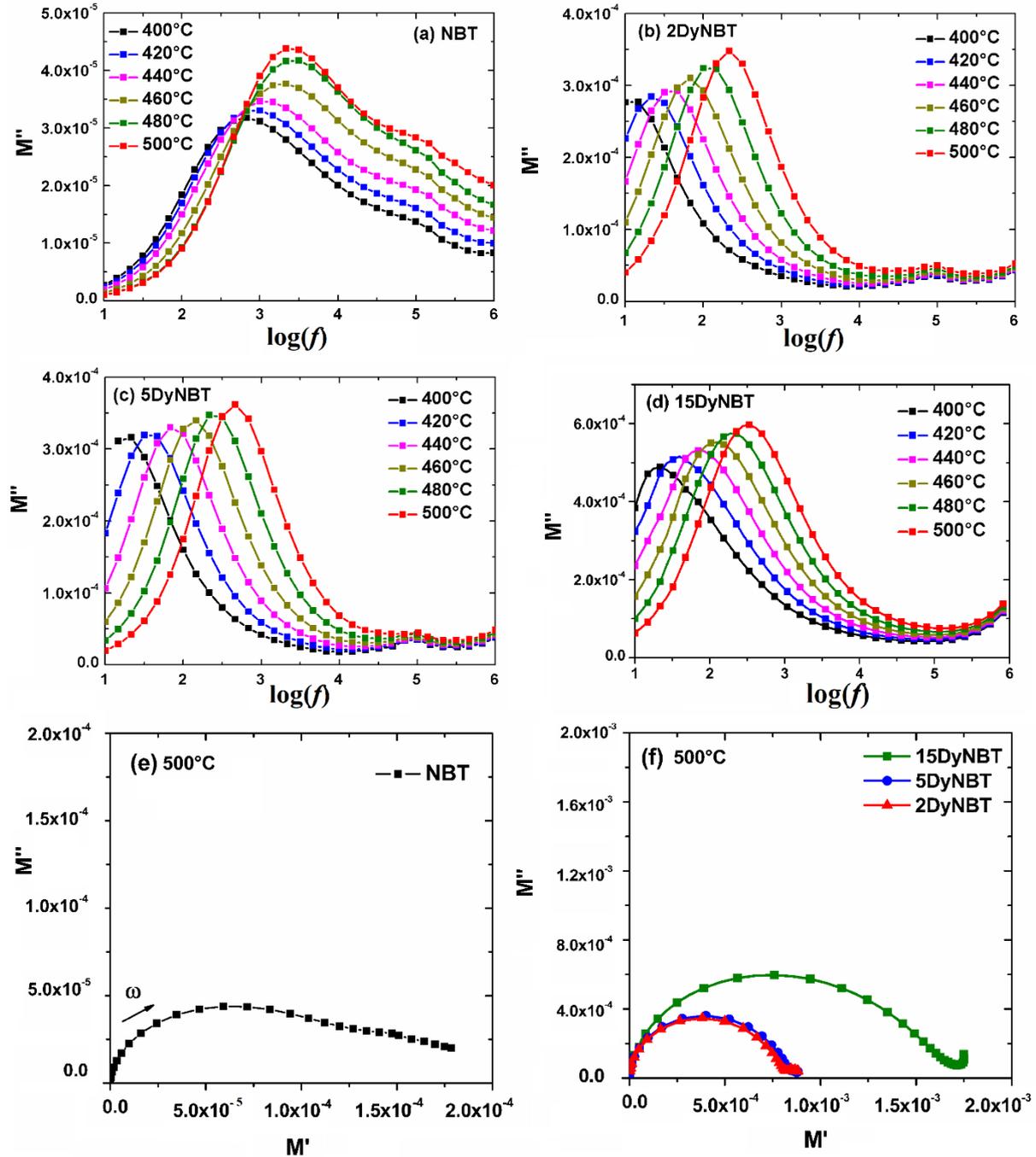

**Figure 4:** Variation of imaginary electric modulus (M″) with frequency of **(a)** NBT, **(b)** 2DyNBT, **(c)** 5DyNBT, and **(d)** 15DyNBT at different temperatures. Nyquist plot of the electric modulus at 500°C for **(e)** NBT, **(f)** doped NBT compounds.

Recall that the imaginary part of the electric modulus (M″) indicates the localized relaxation process whereas the imaginary part of the impedance (Z″) gives an overview of the resistive properties of the material [34]. Moreover, East *et al.* [33] showed up that the electric modulus gives indications of the electrical homogeneity of the bulk component. In fact, for a well sintered ceramic being electrically homogeneous a single peak is observed in M″ with a half-height peak closer to the ideal Debye case (1.14 decades). On the other hand, when the sample is electrically inhomogeneous, two overlapping peaks or a single asymmetric peak, having a half-height peak beyond 1.14 decades, can appear in M″ [33]. Figure 4 (a-d) present the variation of the imaginary part of the electric modulus (M″) as a function of frequency at different temperatures. For the compositions with x = 0, 2, and 5%, two distinguished peaks are observed. These peaks can be attributed to the presence of both grains and grain boundaries effects. In contrary to the other compounds, the small peak at high frequency seems to disappear for the composition with x=15% (figure 4 (d)). Nothing that when the resistance of grains and grain boundaries are of similar order, it is hard to separate both contributions in the impedance plot. Knowing that the capacitance of grains is very low (~pF), we surmise that the high frequency peak is attributed to the grains (~$10^5$ Hz). Therefore, the higher peak being at low frequency can be attributed to the grain boundaries. This is due to the fact that the imaginary part of the electric modulus highlights the contribution having the smallest capacitance value. Noting that the low frequency region of the M″(*f*) curves (below the peak maxima) corresponds to long-range mobility of charge carriers. With frequency increasing in the region above the peak maxima, charge carriers have short-range mobility. Thus, the peak maxima show the change from region where ions have the ability to move over long distances to the region where ions are confined in their potential wall [35]. Figures 4 (e) and (f) show Nyquist plots of M(*f*) at 500°C of all compounds. We observe two confused semicircles for the pure matrix. In contrast, two semicircles separated from each other are observed for 2DyNBT and 5DyNBT samples.

Figure 5 (a-d) presents a comparison of the imaginary complex impedance (Z″) and the electric modulus (M″) as a function of frequency for the pure and doped matrix. Generally, the overlapping of the peak positions of Z″ and M″ curves is a proof of the delocalized or long range relaxation [36]. Our studied systems present a difference in symmetry and peak position in Z″ and M″ curves. For the pure matrix (figure 5 (a)), a clear difference is observed in the peak positions of both curves which confirms a short-range relaxation in NBT. On the contrary, the low doped systems (figure 5 (b and c)) present very close peak positions. In this case, we suggest that both localized and long-range relaxation coexist, which may be due to the introduction of a small amount of the $4f^{10}$ orbital of dysprosium which is known to delocalize charge carriers. Interestingly, we observe that with further doping (x = 15%) the difference between the peaks is even more pronounced (figure 5 (d)), indicating a short-range relaxation for the higher composition.

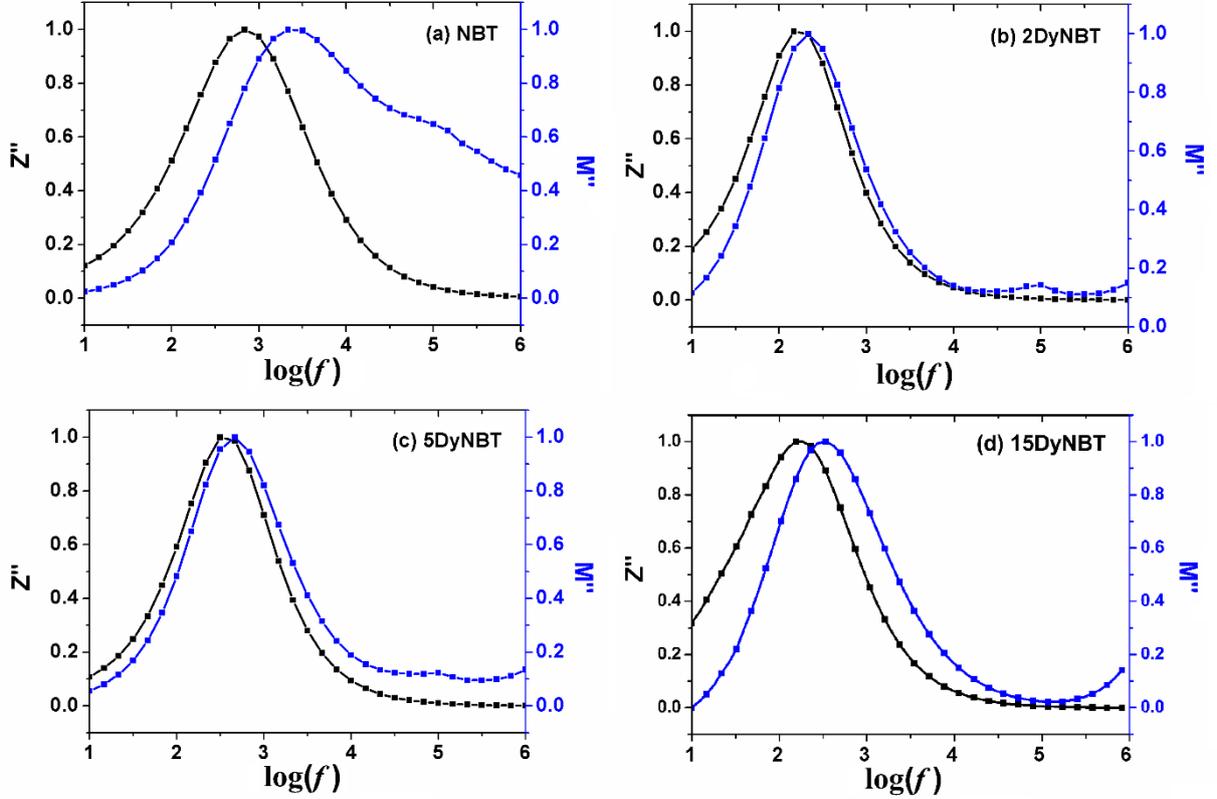

**Figure 5:** Imaginary part of complex impedance (Z″) and electric modulus (M″) as a function of frequency of **(a)** NBT, **(b)** 2DyNBT, **(c)** 5DyNBT, **(d)** and 15DyNBT at 500°C.

Figure 6 (a-d) presents the complex impedance plot (Nyquist plot) of the pure NBT and doped compositions from 400°C to 500°C. We notice that for the pure compound, the curves show well resolved semicircles for all the temperatures. In the case of doping (figure 6 (b, c, d)), we observe incomplete semicircular arcs for temperatures below 460°C. However, with increasing the temperature, the radius of the semicircles decreases forming well resolved semicircles which reflects the semiconducting nature of our systems. The *dc* resistance can be obtained directly from the intersection of the low frequency region with the real axis (Z′). Therefore, with increasing the temperature the *dc* resistance of the material decreases which confirms the NTCR behavior (Negative Temperature Coefficient of Resistivity).

As we have observed in figure 4, both grain and grain boundary effect are present in our studied systems. Therefore, a well exploitation of the Nyquist plot will permit us to examine the number and nature of the semicircles in the Cole-Cole plot, via modelling an equivalent circuit containing, (i) grain (bulk) resistance and capacitance ($R_g$, $C_g$), (ii) grain boundary (interface) resistance and capacitance ($R_{gb}$, $C_{gb}$), in parallel and series combinations ($R_g < R_{gb}$) (insert of figure 6 (e)). Furthermore, the absence of the overlapping peak positions of Z″ and M″ in figure 4 (a-d) implies the departure from ideal Debye behavior and suggests the use of a constant phase element (CPE) in fitting of the circuit [30].

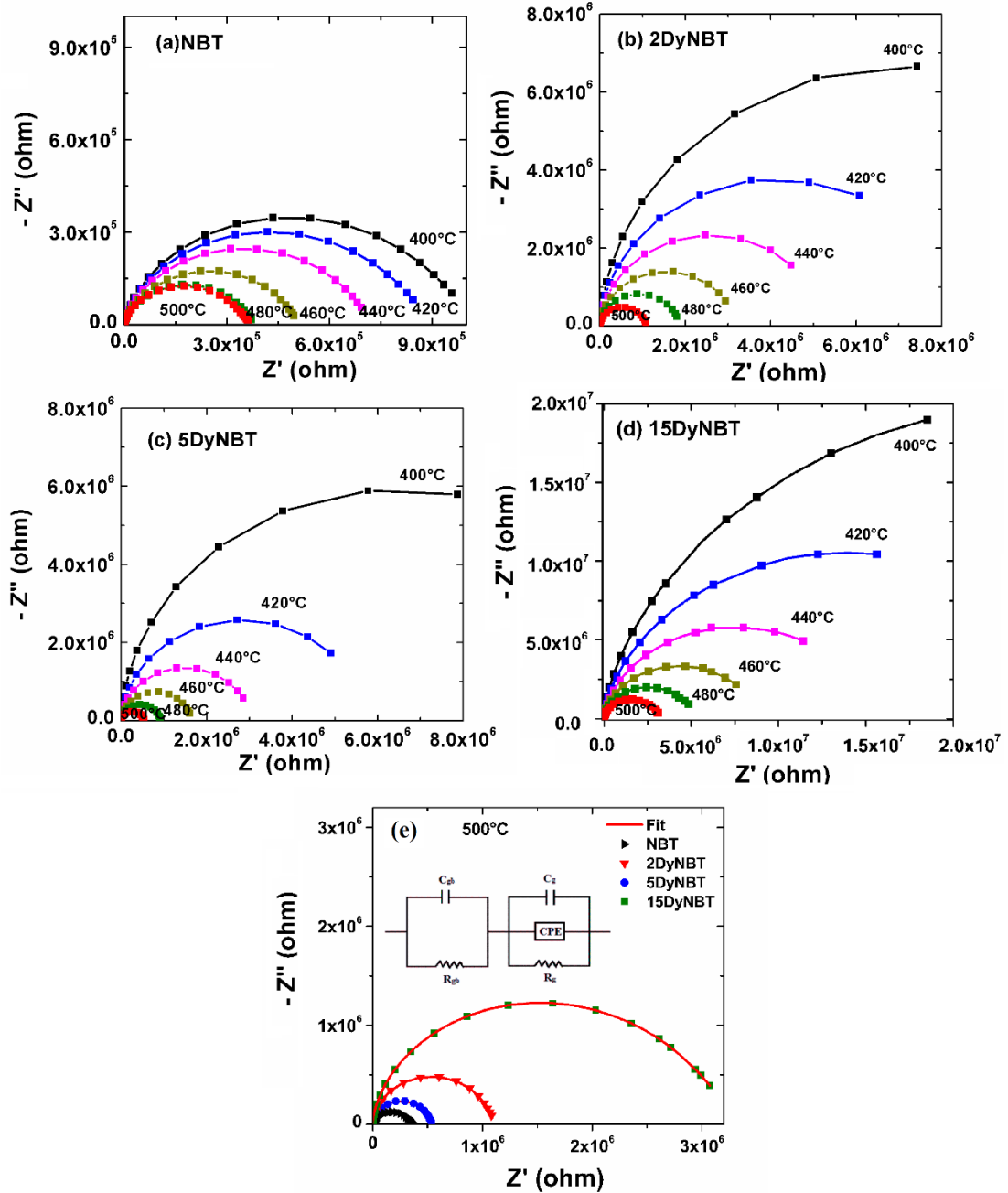

**Figure 6:** Complex impedance plot of **(a)** NBT, **(b)** 2DyNBT, **(c)** 5DyNBT, and **(d)** 15DyNBT at different temperatures, **(e)** the fitted cole-cole plots, inset shows the equivalent circuit of NBT based system.

The CPE admittance is expressed by: $Y(CPE) = A_0(j\omega)^n = A\omega^n + jB\omega^n$ ($A = A_0 \cos(n\pi/2)$ and $B = A_0 \sin(n\pi/2)$), where $A_0$ and n are temperature dependent parameters. An ideal capacitor is described by n = 1, and an ideal resistor is described by n = 0. Figure 6 (e) presents the experimental Cole-Cole plots of the studied compounds at 500°C. This plot permits us to evaluate the effect of doping in NBT matrix on the resistive properties of our systems. By the use of the equivalent circuit (inset of figure 6 (e)), we were able to fit the experimental data of the cole-cole plots at 500°C for all studied compounds. The fitted data are in good agreement with the measured ones. Table 4 groups the fitted resistance and capacitance of grains ($R_g$, $C_g$) and of grain boundaries ($R_{gb}$, $C_{gb}$) for all compounds. Notice that the grain boundary resistance is increasing as a function of doping. This behavior has been reported by Christie *et al.* and Badwal *et al.* for rare earth doped perovskites [37,38]. This can be explained by the decrease of grain size caused by dysprosium introduction. Detailed SEM images can be found in reference [12]. The decrease of grain size will lead to an increase of grain boundaries amount. Consequently, the grain boundary resistance increases due to the reinforced grain boundary-space charge effect. Concerning the grain resistance, we observe that the dysprosium doping increases the resistivity compared to the pure NBT with a highest resistivity given for 15DyNBT sample with a *Pnma* structure. However, a small dysprosium introduction (2DyNBT) resulted in a high resistance compared to the intermediate concentration (5DyNBT). The same behavior has been observed by Khatua *et al*. for erbium doped $0.94Na_{0.5}B_{0.5}TiO_3$-$0.06BaTiO_3$ [39], as well as for Niobium doped $(Na_{0.85}K_{0.15})Bi_{0.5}TiO_3$ [40]. Authors observed that at a certain amount of rare earth (RE) doping, a lower resistivity value is obtained. They correlated this behavior to the phase segregation namely pure NBT and Lanthanide saturated phase regions. So, we infer that in the present work the observation of lower resistivity for the intermediate concentration compared to 2DyNBT composition, could be related to a possible phase segregation not observable by x-ray diffraction technique. Further supporting evidence, such high-resolution SEM, is needed to confirm this assumption. We observe that the grain and grain boundaries resistance for the highest composition (15DyNBT) are of similar order, which explains the difficulty to differentiate between both contributions in figure 4.

**Table 4:** Comparison of electrical parameters $R_{gb}$ (ohm), $C_g$ (F), $R_{gb}$ (ohm), $C_{gb}$ (F) resulting from the fitting of the experimental data using equivalent circuit of xDyNBT at 500°C.

| Composition | $R_{gb}$ (ohm) | $C_{gb}$ (F) | $R_g$ (ohm) | $C_g$ (F) |
|---|---|---|---|---|
| **NBT** | $2.89 \times 10^4$ | $3 \times 10^{-9}$ | $3.32 \times 10^5$ | $2.05 \times 10^{-10}$ |
| **2DyNBT** | $3.2 \times 10^4$ | $3.37 \times 10^{-8}$ | $1.08 \times 10^6$ | $7.08 \times 10^{-10}$ |
| **5DyNBT** | $8.18 \times 10^4$ | $1.07 \times 10^{-8}$ | $4.44 \times 10^5$ | $6.28 \times 10^{-10}$ |
| **15DyNBT** | $1.67 \times 10^6$ | $1.96 \times 10^{-10}$ | $1.53 \times 10^6$ | $2.23 \times 10^{-10}$ |

*3.4. Raman spectroscopy investigations*

*3.4.1. Raman mode analysis*

As mentioned before, Raman spectroscopy is a sensitive method to probe structural transformations, as well as the study of the temperature dependence of phonon modes. Recall that, NBT crystallizes in the rhombohedral *R3c* phase from low temperature to ~200°C. Therefore, a total of 13 modes are Raman active as expressed by the irreducible equation: $\Gamma_{Raman,R3c} = 4A_1 + 9E$. $A_1$ and E symmetry being the Raman modes associated with lattice displacements parallel and perpendicular to the c-axis of the unit cell, respectively [41]. Figure 7 presents the ambient-temperature Raman spectra of xDyNBT ceramics including a deconvolution of Raman spectrum of NBT. The sodium bismuth titanate Raman response is seen to be comparable with previous reports [42–47]. Notice that our spectrum includes low frequency Raman part <100 cm$^{-1}$. The presented spectra show the existence of a low frequency mode around 50 cm$^{-1}$. Numerous studies didn't include this last frequency part while studying NBT Raman spectra, despite of its remarkable importance. In fact, while studying the doping effect of dysprosium into the A-site of $Na_{0.5}Bi_{0.5}TiO_3$ system, it is important to highlight its impact on the A-site cationic vibrations found in the low frequency part. However, some authors were particularly interested in this region [48–51]. For instance, Siny *et al.* evaluated the behavior of the low frequency mode to highlight the existence of a central component in NBT single crystal, known in relaxor ferroelectrics [51,52]. Other studies of NBT-xBT single crystal [48] as well as ceramic [49] systems, have investigated the temperature evolution of the Raman spectra of the low frequency mode in order to probe the doping effect of $BaTiO_3$ into NBT system.

As it can be seen in figure 7, the deconvolution of the Raman spectrum of NBT includes seven Raman modes. For a better following of NBT Raman modes, we designate each mode with $\nu_i$ where 'i' is the mode number, from low to high frequency. The first low frequency mode $\nu_1$ ~55 cm$^{-1}$, is assigned to phonon modes localized in the A-site cations [53]. This mode is sensitive to the A-site cationic off-centering, where A-site cations in xDyNBT system involves $Na^+$, $Bi^{3+}$ and $Dy^{3+}$ ions. However, a reverse Monte Carlo simulation on neutron total elastic scattering (TES) of pure NBT system, demonstrates that $Bi^{3+}$ have a higher affinity toward off-centering from the twelve surrounding $O^{2-}$ anions compared to $Na^+$. Such difference is due to the different Na-O (ionic bonding) and Bi-O (mixed ionic-covalent bonding) interactions [54]. One can say that $Bi^{3+}$ off-centering is larger and less dispersed than that of $Na^+$ cations. Thus, $\nu_1$ will probably evidence Bi-O/Dy-O vibrations instead of Na-O ones. This last is observed at ~55 cm$^-$ for pure NBT and seems to slightly hardens in frequency while doping by Dy element, which can be explained by the introduction of the lower cationic mass. In addition, an evident broadening is clearly observed with dysprosium introduction, which is related to the disorder resulting from the lower ionic radii of $Dy^{3+}$ (0.912 Å) compared to $Bi^{3+}$ (1.03 Å).

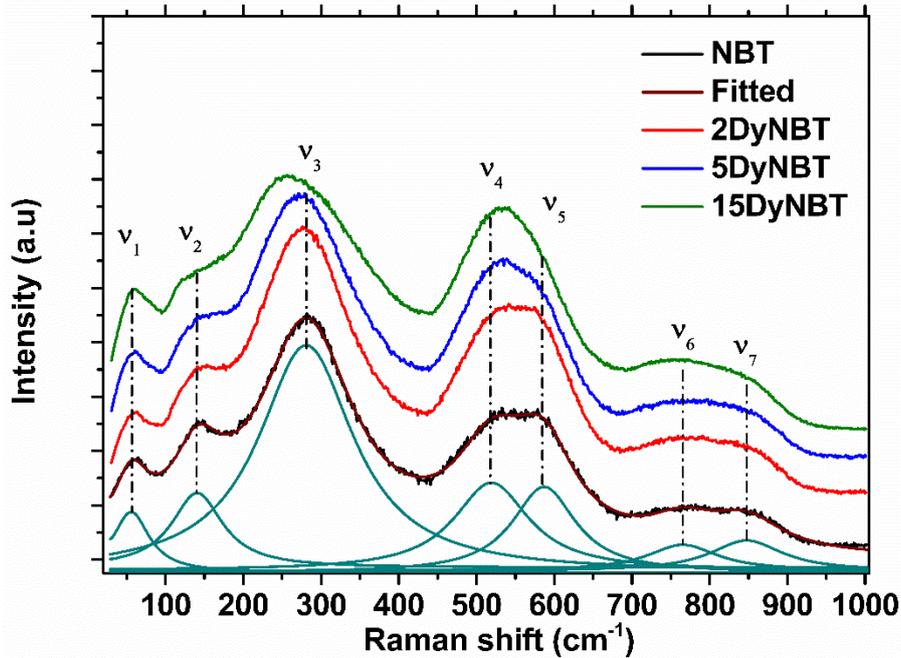

**Figure 7:** Raman spectra at room temperature of xDyNBT (x = 0, 2, 5, and 15%) ceramics. The deconvolution of Raman spectrum of pure NBT is added showing the presence of seven modes.

Regarding the $A_1$ symmetry mode $\nu_2$ ~150 cm$^{-1}$, it is also assigned to A-site cationic vibrations [55,56]. However, by considering the cationic mass, Na-O vibrations will be assigned to $\nu_2$. G. de la Flor *et al.* considered this vibrational mode as an A-BO$_3$ translation mode involving both A-site and B-site cation vibrations for NBT-xBT ceramics, and consequently it is the hearth of coupling processes between A-site and B-site off-centering [48]. Moreover, the dominating feature in our spectra is seen to be the broad and intense vibrational Raman mode $\nu_3$~280 cm$^{-1}$ mainly involving B-site cations, thenceforth Ti-O vibrations [56]. As it can be seen from the plots, an apparent downshift in frequency is observed for $\nu_3$ as a function of doping. Kreisel *et al.* have assigned this mode as being an $A_1$ mode related to the strength of the Ti-O bond [57]. It seems however, that Ti-O bond length became longer with the introduction of dysprosium, therefore the weakening of its bond-stretching forces. Due to the small ionic radius of $Dy^{3+}$ compared to $Bi^{3+}$, the TiO$_6$ octahedra have undergone high distortions for the high concentrations (15DyNBT) of dysprosium which resulted in a symmetry change to an orthorhombic *Pnma* structure, with the appearance of a new mode around 300 cm$^{-1}$.

Furthermore, the enlargement of $\nu_1$, $\nu_2$, and $\nu_3$ as a function of $Dy^{3+}$ could be a signature of a local disorder within the matrix. Concerning the high frequency modes $\nu_4$ ~519 cm$^{-1}$, $\nu_5$ ~586 cm$^{-1}$, $\nu_6$ ~765 cm$^{-1}$ and $\nu_7$ ~847 cm$^{-1}$, they are all assigned to BO$_6$ tilting modes [56], and thus are resulting from TiO$_6$ octahedral vibrations. These modes are well known in perovskites to involve oxygen atoms vibrations, while cations are almost at rest [58]. In the present study, doping with $Dy^{3+}$ leads to a frequency downshift of $\nu_4$ and $\nu_5$ modes, but the vibrational modes $\nu_6$ and $\nu_7$, situated at the higher frequencies, doesn't undergo any perceptible variation in frequency. Note that the ratio intensity of $\nu_4$ and $\nu_5$ (I($\nu_4$)/ I($\nu_5$)) is observed to increase with higher $Dy^{3+}$ concentrations. For the highest concentration, $\nu_5$ mode seem to disappear due to the symmetry change.

Figure 8 (a-c), presents the depolarized Bose-Einstein corrected Raman spectra of pure NBT, 2DyNBT and 5DyNBT ceramics (all these compounds present a polar rhombohedral structure at room temperature) from room temperature to 580°C. The persistence of Raman peaks at elevated temperatures is evidenced from the plots. Keep in mind that NBT has a cubic paraelectric (*Pm3-m*) structure above 520°C for which no Raman active modes are allowed. However, based on neutron total elastic scattering (TES) [59] and x-ray absorption spectroscopy (XAS) [60] studies the persistence of $\nu_1$ and $\nu_2$ modes at higher temperatures indicates the presence of A-site off-centering cations. In the same way, the persistence of $\nu_3$ and the high frequency modes indicates B-site cations off-centering, and ferroically deformed BO$_6$ octahedra, respectively even in the paraelectric state.

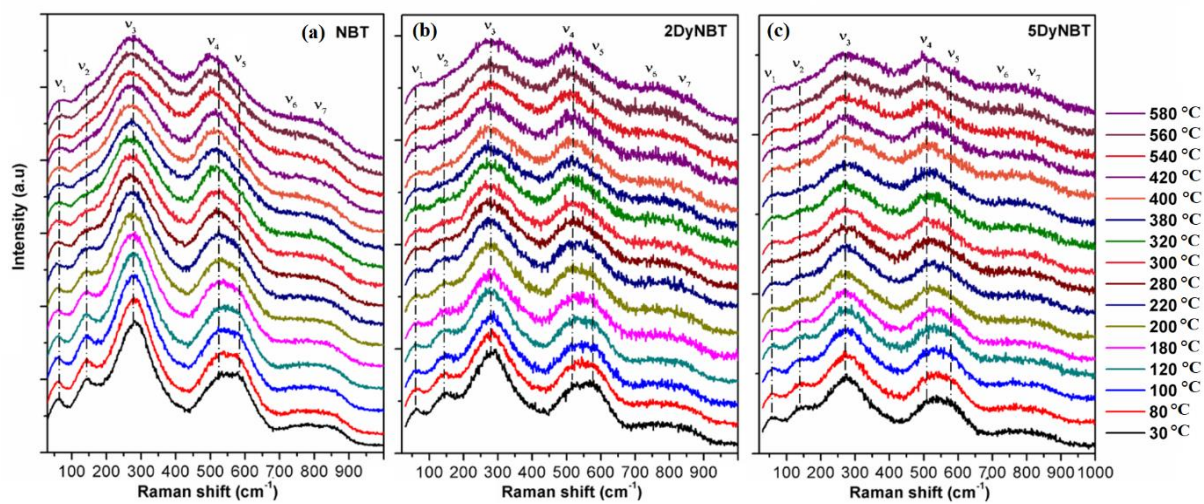

**Figure 8:** Temperature dependent Raman spectra of **(a)** NBT, **(b)** 2DyNBT, and **(c)** 5DyNBT compositions from 30°C to 580°C, dot lines are guide for the eyes.

As it can be depicted from figure 8, for NBT, $\nu_1$ (Bi-O vibrational mode) is slightly shifting to lower frequencies as the temperature increases accompanied with a clear band enlargement.

However, for 2DyNBT and 5DyNBT compositions, a continuous hardening is observed for $\nu_1$. We note herein that above ~440°C, $\nu_2$ almost vanished for the doped compositions. Conversely to the pure compound where $\nu_2$ resists until 580°C. On the other hand, a subtle downshift is observed in $\nu_3$ for xDyNBT compounds with a clear intensity decrease and broadening as the temperature increases. Worthwhile, $\nu_4$ is the mode that gives a significant comparison between the studied compounds. Indeed, a significant $\nu_4$- downshift is observed as a function of temperature for NBT. However, this downshift in frequency is slowed by the incorporation of dysprosium, then stopped with increasing doping concentration. That is the case of the composition with 5DyNBT doping where almost no frequency downshift could be detected while the temperature increased for $\nu_4$. Furthermore, $\nu_5$ is observed to vanish for all three composition for high temperatures > ~320°C. No distinct changes are reported for the higher frequencies Raman modes $\nu_6$ and $\nu_7$.

### 3.4.2. Raman approach of phase transitions

Note that, Na/Bi/Dy occupies the 2a (x,x,x) positions in the rhombohedral *R3c* structure, and are seen to displace along the $[111]_p$ direction. Whereas in the orthorhombic *Pnma* structure, Na/Bi/Dy occupies the 4c (x,1/4,z) positions are showing antiparallel displacement along $[100]_p$ direction [61]. Therefore, vibrational modes associated with A-site cations are expected to be sensitive toward structural changes. Besides, Ti-O bonds are also responsible for driving ferroelectricity in the system, it's hence suggested that this particular phonon can be directly involved in structural phase transitions. Thus, we have undertaken in this part of work, a deep analysis of the thermal evolution of $\nu_1$, $\nu_2$, and $\nu_3$ vibrational Raman modes for the three compounds. For this purpose, the FWHM ($\Gamma$), the wavenumber ($\omega$), and the intensity (I) of $\nu_1$, $\nu_2$, and $\nu_3$ modes for our studied compounds have been followed as a function of the temperature.

The temperature dependencies of the wavenumber of $\nu_2$ ($\omega_2$) and the wavenumber of $\nu_3$ ($\omega_3$) for the three compounds are depicted in figure 9. We can clearly observe a behavior change ~320°C for NBT and ~340°C for 2DyNBT which matches well with their related $T_m$ regions. Curiously, for 5DyNBT the minimum of $\omega_2$ and $\omega_3$ is seen at ~200°C. Remind that, the depolarization temperature ($T_d$ ~200°C) seen in figure 2, can be observed easily and is more pronounced for the doped compositions. Especially for 5DyNBT, where the enlargement and the decrease in the dielectric permittivity values at $T_m$ permits to $T_d$ anomaly to become more pronounced than the very broad high temperature anomaly at $T_m$. As mentioned above, the AFE behavior is more sensitive to be detected by Raman spectroscopy instead of conventional x-ray diffraction.

Thus, one can conclude that the anomaly observed ~200°C may correspond to the depolarization temperature. This behavior is an indication of the pronounced depolarization temperature compared to the broad $T_m$ anomaly for the 5DyNBT compound observed in the dielectric properties. Concerning the intensity of $v_2$ ($I_2$) and the FWHM of $v_3$ ($\Gamma_3$) (Figure S2 and S3 in the supplementary material), a behavior change from the linearity occurs near $T_m$ region for the three studied compounds.

While following the thermal evolution of the FWHM ($\Gamma$) of $v_1$ for the three compounds (see figure S2 in the supplementary material), we noticed maxima near the orthorhombic-to-tetragonal phase transition ($T_m$). Another interesting behavior have been observed in the study of $\Gamma$, which is the appearance of a hump near the depolarization temperature ($T_d$) describing the FE-to-AFE like phase transition. Interestingly, the hump observed in the depolarization region of 5DyNBT is prominent compared to that of NBT. The observation of this pronounced hump near $T_d$ for 5DyNBT, corroborates the dielectric investigations, where $T_d$ becomes prominent and veil the broad $T_m$ (Figure 1).

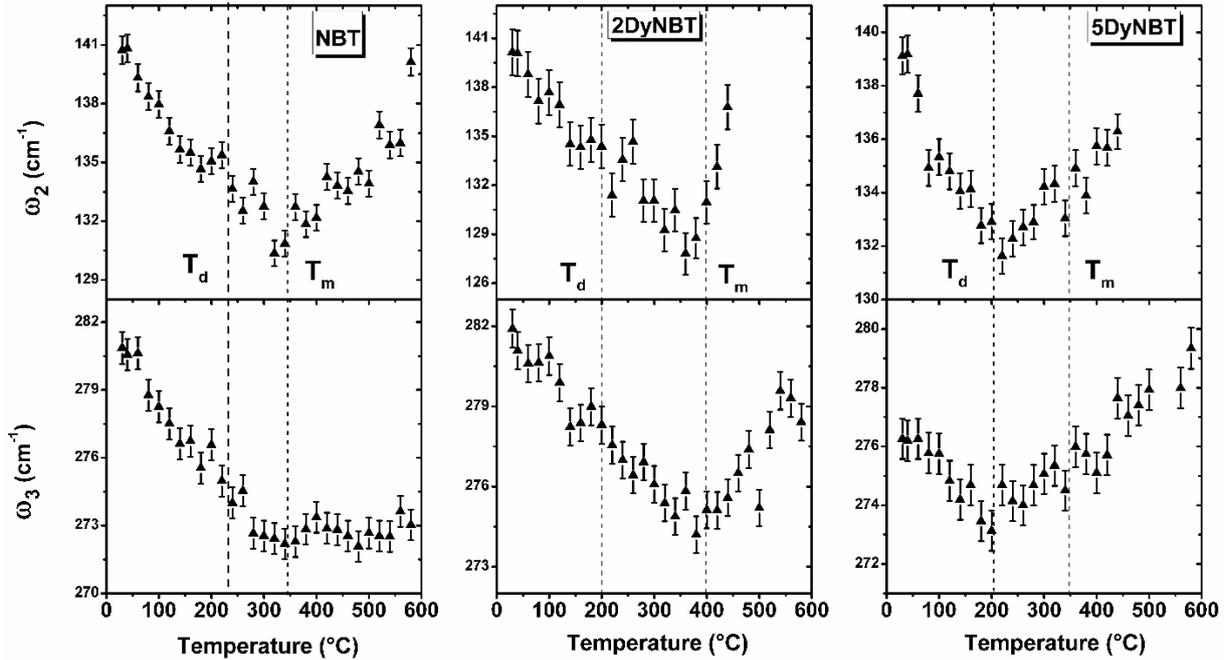

**Figure 9:** Temperature evolution of the wavenumber $\omega_2$ and $\omega_3$ for NBT, 2DyNBT, and 5DyNBT compositions from 30°C to 580°C. As a guide for the eyes, we added in figures the regions of $T_d$ and $T_m$ extracted from the dielectric investigations.

## 4. Conclusion

$Na_{0.5}(Bi_{1-x}Dy_x)_{0.5}TiO_3$ ceramics system have been studied. The structural refinement showed that the incorporation of $Dy^{3+}$ until 5% seems not to change the structural chemistry. However, for 15% of $Dy^{3+}$ introduction, the symmetry changes from rhombohedral *R3c* to orthorhombic *Pnma* structure. Dielectric investigations are found to highlight both the ferroelectric to antiferroelectric like ($T_d$), and antiferroelectric to paraelectric ($T_m$) phase transitions. We report that as a function of dysprosium introduction, $T_m$ is decreasing in intensity while $T_d$ is getting more pronounced. Complex impedance spectroscopy was performed in a wide frequency range ($10 - 10^6$ Hz) at different temperatures for a better understanding of the electrical properties and the correlation with the microstructure of the materials. Non-Debye process was found on our samples with a FWHM superior to 1.14 decades. Moreover, NBT was found to be characterized by a short-range relaxation. Contrary to the doped xDyNBT (x=2 and 5%) compositions were both localized and long-range relaxation of charge carriers coexist. Concerning the high concentration 15DyNBT, we highlighted a short-range relaxation. The cole-cole plots are fitted with an equivalent circuit based on the brick-layer model in order to extract the resistance and capacitance of grains and grain boundaries of our samples. We found out that with dysprosium introduction, the grain boundary resistance is increasing as a result of the reinforced grain boundary-space charge effect. Whereas the grain resistance is seen to have the highest value for 15DyNBT composition. In situ Raman spectroscopy have been conducted on the ferroelectric compounds, over a wide temperature range from room temperature to 580°C and detailed analysis of the different vibrational modes are performed. Correlation between the phase temperature transitions in the dielectric measurements and the temperature-induced transformation processes is discussed. In fact, the temperature evolution of the low frequency modes for NBT and 2DyNBT show the appearance of an anomaly in $T_m$ temperature range in agreement with dielectric measurements. Conversely, the anomaly seen in 5DyNBT is found in the temperature region corresponding to the depolarization temperature. We directly correlated this behavior with the effect of doping on the dielectric properties of our samples, where $T_d$ is prominent compared to the broad $T_m$ for this last composition.

## Acknowledgments

Financial support by the Haute France Region/ FEDER (project MASENE) and H2020-RISE-ENGIMA project are gratefully acknowledged.

## Tables and figures captions

**Table 1:** Details of Rietveld refinement parameters for NBT, 2DyNBT, 5DyNBT and 15DyNBT compositions.

**Table 2:** Refined structural parameters of NBT, 2DyNBT, 5DyNBT and 15DyNBT compositions.

**Table 3:** Selected inter-atomic distances (Å) and O-Ti-O angles for NBT, 2DyNBT, 5DyNBT and 15DyNBT compositions.

**Table 4:** Comparison of electrical parameters $R_{gb}$ (ohm), $C_g$ (F), $R_{gb}$ (ohm), $C_{gb}$ (F) resulting from the fitting of the experimental data using equivalent circuit of xDyNBT at 500°C.

**Figure 1**: **(a)** X-ray diffraction patterns of all xDyNBT (x=0, 2, 5 and 15%) sintered ceramics. Rietveld refinement X-ray powder diffraction pattern of **(b)** 5DyNBT and **(c)** 15DyNBT: experimental, calculated and their difference. The projection views of the structures are shown in the inset.

**Figure 2:** Dielectric constant of **(a)** NBT, **(b)** 2DyNBT, **(c)** 5DyNBT, and **(d)** 15DyNBT at three different frequencies. **(e)** Dielectric losses at 100 kHz.

**Figure 3:** Variation of the imaginary part of complex impedance (Z″) with frequency **(a)** NBT, **(b)** 2DyNBT, **(c)** 5DyNBT, **(d)** 15DyNBT at different temperatures, and **(e)** variation of the real part of complex impedance (Z′) with frequency for all compounds at 500°C.

**Figure 4:** Variation of imaginary electric modulus (M″) with frequency of **(a)** NBT, **(b)** 2DyNBT, **(c)** 5DyNBT, and **(d)** 15DyNBT at different temperatures. Nyquist plot of the electric modulus at 500°C for **(e)** NBT, **(f)** doped NBT compounds.

**Figure 5:** Imaginary part of complex impedance (Z″) and electric modulus (M″) as a function of frequency of **(a)** NBT, **(b)** 2DyNBT, **(c)** 5DyNBT, **(d)** and 15DyNBT at 500°C.

**Figure 6:** Complex impedance plot of **(a)** NBT, **(b)** 2DyNBT, **(c)** 5DyNBT, and **(d)** 15DyNBT at different temperatures, **(e)** the fitted cole-cole plots, inset shows the equivalent circuit of NBT based system.

fitting of the experimental data using equivalent circuit of xDyNBT at 500°C.

**Figure 7:** Raman spectra at room temperature of xDyNBT (x = 0, 2, 5, and 15%) ceramics. The deconvolution of Raman spectrum of pure NBT is added showing the presence of seven modes.

**Figure 8:** Temperature dependent Raman spectra of **(a)** NBT, **(b)** 2DyNBT, and **(c)** 5DyNBT compositions from 30°C to 580°C, dot lines are guide for the eyes.

**Figure 9:** Temperature evolution of the wavenumber $\omega_2$ and $\omega_3$ for NBT, 2DyNBT, and 5DyNBT compositions from 30°C to 580°C. As a guide for the eyes, we added in figures the regions of $T_d$ and $T_m$ extracted from the dielectric investigations.